\documentclass[aps,showpacs,floats,superscriptaddress]{revtex4-2}

\usepackage{graphicx}
\usepackage{epstopdf}
\usepackage{amssymb}
\usepackage{amsmath,bm}
\usepackage{psfrag}
\usepackage{epsfig}
\usepackage{float}
\usepackage{bm}
\usepackage{textcomp}
\usepackage{hyperref}

\begin{document}
\title{On Nanowire Morphological Instability and Pinch-Off by Surface Electromigration}

\author{Mikhail Khenner\footnote{Corresponding
author. E-mail: mikhail.khenner@wku.edu.}}
\affiliation{Department of Mathematics, Western Kentucky University, Bowling Green, KY 42101, USA}
\affiliation{Applied Physics Institute, Western Kentucky University, Bowling Green, KY 42101, USA}

\begin{abstract}
\noindent
Surface diffusion and surface electromigration may lead to a morphological instability of thin solid films and nanowires.
In this paper two nonlinear analyses
of a morphological instability are developed for a single-crystal cylindrical 
nanowire that is subjected to an axial current. These treatments extend the conventional linear stability analyses without surface electromigration, 
that manifest a Rayleigh-Plateau instability.
A weakly nonlinear analysis is done slightly above the Rayleigh-Plateau (longwave) instability threshold. It results in a one-dimensional Sivashinsky 
amplitude equation that describes a blow-up of a surface perturbation amplitude 
in a finite time. 
This is a signature of a pinching singularity of a cylinder radius, which leads to a wire separation into a disjoint 
segments.
The time- and electric field-dependent dimensions 
of the focusing self-similar amplitude profile approaching a blow-up are characterized via the scaling analysis.
Also, a weakly nonlinear multi-scale analysis is done at the arbitrary distance above a longwave or a shortwave instability threshold.
The time- and electric field-dependent Fourier amplitudes of the major instability 
modes are derived and characterized.

\end{abstract}

\date{\today}
\maketitle


\section{Introduction}
\label{Intro}

Surface electromigration \cite{Huntington,HoKwok,REW} is a well-known and efficient method to guide morphological changes of thin films by surface 
diffusion. In particular, it has been used to manufacture nanocontacts by breaking of thin films and nanowires via a controlled pinch-offs 
\cite{PLAPM,VFDMSKBM,AGLCH}. Such nanocontacts are required for engineering and biomedical applications, for example, 
for measurement of the electrical conductance and electronic properties of a molecule, but forming a quality nanocontact is still a major challenge \cite{H}.
To overcome that challenge, the first step should be a basic understanding of a surface electromigration-driven morphological instability 
of a single-crystal wire that leads to a pinch-off event. 

Modeling morphological instabilities of nanowires by a classical approach of accounting for surface diffusion via an evolution partial 
differential equation (PDE) for a shape variable \cite{NM2,NM1,C} flourished in the 1990s and early 2000s \cite{BBW,CFM,McCVoorhees,WMVD,KDMV,GM,GMM}. 
A recent paper emphasizing this approach is by Wang \textit{et al.} \cite{WTLZNSMH}. 
These authors considered a surface area minimization problem, computed a wire pinch-off using a phase-field method,  and extended the Rayleigh-Plateau stability condition to finite amplitude perturbations. 
Also limited Monte Carlo computations were published \cite{GP}. 
For predictive applied modeling of a nanocontact fabrication process a multi-physics framework is needed, that explicitly includes 
a model and a computation of a nanowire pinch-off instability, whereby the latter is triggered and enhanced by the surface current. Since the cited works do not consider the key mass transport mechanism, 
namely surface electromigration (shortened to electromigration in the rest of the paper), they cannot form a basis for that framework. 

Toward the stated goal, in Ref. \cite{MyWire1} this author introduced a PDE-based model for electromigrated cylindrical nanowire deposited on a substrate. That model is complicated by the presence of 
the contact lines and a non-axisymmetric surface instability modes. A linear stability analysis (LSA) of that model is performed in Ref. \cite{MyWire1}, from which a simpler
case of axisymmetrically evolving free-standing wire is easily recovered. Axisymmetric modes dominate evolution of surface perturbations for a free-standing wire in the absence of 
electromigration, i.e. when only a natural high-temperature surface diffusion is operative \cite{BBW}. 
Provided the axial current and a corresponding diffusion anisotropy that is a function of the axial variable, this is expected to hold when the electromigration is also operative.
Ref. \cite{MyWire2} reports a computation of a wire breakup into a chain set of particles for the simpler case. 
The breakup time and the number of particles that emerge upon a breakup are characterized as a function of the initial surface roughness.

In this communication, also assuming axisymmetrically evolving, free-standing long wire, in Sec. \ref{Weak} I perform a 
weakly nonlinear analysis 
slightly above the longwave instability threshold.  The latter means that a critical (a most dangerous, i.e. a 
fastest growing) sinusoidal perturbation of a uniform surface profile $r(y)=const.$ (where $r$ is a wire radius and $y$ the 
axial coordinate) 
has a tiny axial wavenumber $k\rightarrow 0$. 
I derive an amplitude equation via the expansion of a governing PDE up to the terms that are quadratic in the amplitude, 
whereas LSA is based on the expansion up to linear terms, i.e. a linearization procedure. 
In other words, the perturbation amplitude is finite in the weakly nonlinear analysis. This key underlying assumption is the same in this paper and 
in Ref. \cite{WTLZNSMH}. A finite amplitude is quite likely to occur in experiment \cite{MSLCM}. A scaling analysis and a computation of the 
amplitude equation show that the amplitude becomes unbounded (blows up) in finite time. Following Ref. \cite{GB} and to simplify writing, 
I will refer to the amplitude approaching a blow-up singularity as a spike. From the analysis I obtain the much needed information on the 
time-and-electric field-dependent dimensions of the spike. 
Note that the blow-up singularity of the amplitude (a spike formation) is the 
signature of a finite-time pinching singularity of a wire radius $(r\rightarrow 0)$. The (finite) time of a singularity formation is the same 
for the amplitude singularity and for the radius singularity. In a basic experiment demonstrating a Rayleigh-Plateau 
instability without electromigration  
a wire pinches off at multiple locations along 
the axis, thus several shorter cylindrical wire segments are formed \cite{XLL,KTBECKN}. The spacing between pinch-offs is approximately equal to the 
most dangerous wavelength.
Depending on a new segment's length (that is, depending on whether a Rayleigh-Plateau instability is again operational) a segment may further 
pinch-off resulting in even shorter segments, 
or, provided sufficient time, it may transform by surface diffusion into a spherical nanoparticle.

Independent of the weakly nonlinear analysis, in Sec. \ref{MultScale} I start with the assumption of the initial unstable surface perturbation having an arbitrary 
finite wavenumber 
$0<k<k_c$, where $k_c$ is the instability cut-off wavenumber from LSA, 
and ask which unstable surface modes are the fastest growing. 
In this weakly nonlinear multi-scale treatment the spike formation is not considered. 
(As in the ``standard" weakly nonlinear analysis in Sec. \ref{Weak}, the governing PDE for the surface variable will be expanded to second order in 
a small parameter that has a meaning of a perturbation amplitude. This
explains why I use the term ``weakly nonlinear multi-scale analysis".)
Nonlinearity of the governing PDE forces a fast distortion of the initial sinusoidal surface perturbation into a general smooth surface shape, 
and as was just explained, I am interested in finding the fastest growing Fourier modes in its spectrum.

\section{The model}
\label{Model}

\begin{figure}[h]
\vspace{-0.2cm}
\centering
\includegraphics[width=2.5in]{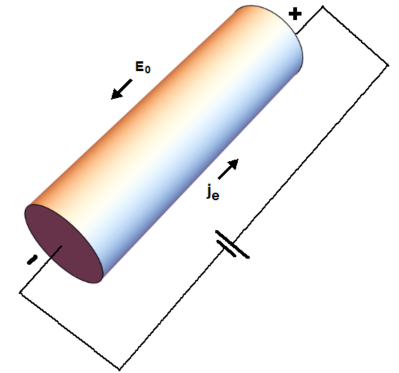}\hspace{0.5cm}
\includegraphics[width=2.5in]{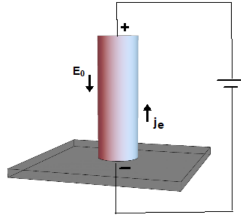}
\caption{Two situations for a free-standing unperturbed cylindrical nanowire (the base state). Left: a nanowire suspended in 3D space, 
say, by 
hanging it on a thread attached to a support. Right: a nanowire grown vertically on a horizontal substrate.  $j_e$ is the surface electric current that drives the surface electromigration of 
adatoms via the "electron wind" \cite{Huntington,HoKwok,REW}.}
\label{Cylinder}
\end{figure}

\begin{figure}[h]
\vspace{-0.2cm}
\centering
\includegraphics[width=2.5in]{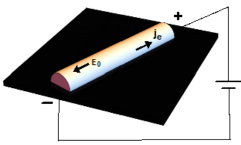}
\caption{Unperturbed cylindrical nanowire grown laterally on a substrate with a 90$^\circ$ contact angle. }
\label{Cylinder_AlongSubstrate}
\end{figure}

Assuming an axisymmetric evolution of a wire surface and a constant electric field $E_0$ in the axial direction, 
I consider surface electromigration on a free-standing cylindrical nanowire of radius $R_0$, see Fig. \ref{Cylinder}. 
A wire material is fcc metal such as gold.
In axisymmetric case the linear stability results in Sec. \ref{LinStab} 
also hold for a wire grown laterally on a substrate - if the base state is semi-cylindrical, meaning that a wire  
surface makes a 90$^\circ$ contact angle with the substrate \cite{MyWire1} (Fig. \ref{Cylinder_AlongSubstrate}). This is because when a 
perturbation is axisymmetric and the contact angle is 90$^\circ$,  a wire does not ``feel" the contact lines in the linear 
approximation. Thus the results of a nonlinear analyses in Sections \ref{Weak} and \ref{MultScale} for a free-standing wire may be considered as approximations
in the case of a laterally grown wire with a 90$^\circ$ contact angle.  

Let $R_0$ and $R_0^4 k T/(D\gamma\Omega^2\nu)$ be the length and the time scales. Here, $k T$ is the thermal energy, 
$\gamma$ the surface energy, $\Omega$ the atomic volume, and $D,\ \nu$ the surface diffusivity and the surface density of the adatoms. 
Then the dimensionless wire radius $r(t,y)$ is governed by a nonlinear evolution PDE \cite{MyWire1,MyWire2}:


\begin{equation}
r_t=\frac{1}{r}\frac{\partial}{\partial y} \left[M(r_y)\frac{r K_y+E}{\sqrt{1+r_y^2}}\right], \label{govPDE}
\end{equation}
where $y$ is the axial variable, 
\begin{equation} 
K=\frac{1}{r\sqrt{1+r_y^2}}-\frac{r_{yy}}{\left(1+r_y^2\right)^{3/2}} \label{curv}
\end{equation}
the mean curvature of the surface, and
\begin{equation}
M(r_y)=\frac{1+S\cos^2{\left[m\left(\arctan{r_y}+\psi\right)\right]}}{1+S}
\label{mobility}
\end{equation}
the anisotropic diffusional mobility of the adsorbed atoms (adatoms) \cite{SK}. Here $S>0$ is the anisotropy strength, $m$ the number of symmetry axes,  
and $\psi$ the misorientation angle, i.e. this is the angle, in the $yz$ plane, formed by a wire axis and the average orientation of the unit normal to the surface.
To reduce the number of parameters in this study it is assumed that a wire axis is along [110] crystalline direction, thus $m=1$
and $0\le \psi \le \pi/2$ \cite{DM}.
Also $E=Q \Delta V R_0^2/(\Omega \gamma \ell)$ is the electric field parameter, where $Q>0$ is the effective charge of ionized adatoms, 
$\Delta V$ a voltage difference between the front and back faces of a wire, $\ell$ the wire length, $\Omega$ the atomic volume, and $\gamma$ the surface energy.
Note that $E_0=\Delta V/\ell$, 
and thus a positivity (a negativity) of $E_0$ and $E$ depends on the sign of $\Delta V$. When surface diffusion is isotropic ($S=0 \rightarrow M(r_y)=1$) 
and the electric field is off ($E=0$), Eq. (\ref{govPDE}) reduces to a well-known evolution equation 
$
r_t=\frac{1}{r} \frac{\partial}{\partial y}\left[\frac{r K_y}{\left(1+r_y^2\right)^{1/2}}\right]
$
\cite{BBW,CFM}, which expresses only the adatom surface diffusion via a surface Laplacian of mean curvature. 
%

The key simplifying physical assumption underlying the above mathematical model is that a current crowding at the tip of a spike is 
neglected.
In the first approximation, a current crowding may be accounted for by replacing $E$ with $E/r$ in Eq. (\ref{govPDE}) \cite{SK,B1,B2}.
This change would result in a complex-valued perturbation growth rate in the linear stability analysis in Sec. \ref{LinStab}, with 
the real part of the growth rate unaffected. Thus an initial surface perturbation would not only grow, but it would also drift 
along a wire, and thus a pinch-off would occur at a location that is different from the location of a minimum of an 
initial perturbation. More important is that neglecting a current crowding means neglecting Joule heat production in a wire, 
which leads to a slower time to a pinch-off - since a higher temperature speeds up the surface diffusion. 
A wire may even melt before it pinched off \cite{H,YS,T}. Faceting of a wire surface may also occur \cite{SSOAUTSSAN} and may 
affect a pinch-off. Lastly, a polycrystal wires are less stable than a single crystal ones due to  
grain-boundary grooving, which may be enhanced by electromigration. It was shown that the spacing between the wire segments 
and the size of the segments are smaller for polycrystal wires \cite{KTEBCKN}.

\section{Linear Stability Analysis}
\label{LinStab}

Eq. (\ref{govPDE}) has the trivial solution $r=1$ (the base state), corresponding to unperturbed cylinder. Introducing a tiny 
perturbation of the base state, $r=1+\xi_0 e^{\sigma t}\cos{ky},\; \xi_0\ll 1$, and linearizing in $\xi_0$ 
gives the growth rate $\sigma$ and the instability cut-off 
wavenumber $k_c$ \cite{MyWire1}:
\begin{equation}
\sigma(k; E)=\frac{2+S(1+\cos{2 \psi})}{2(1+S)}k^2\left(1-k^2\right)+E \frac{S \sin{2 \psi}}{1+S}k^2\equiv 
\frac{\alpha_1+\alpha_3}{2\left(\alpha_3-1\right)}k^2\left(1-k^2\right)+\frac{\alpha_2}{\alpha_3-1}Ek^2,
\label{sigma_axi}
\end{equation}
\begin{equation}
k_c(E)=\sqrt{1+\frac{2 S E \sin{2 \psi}}{2+S(1+\cos{2 \psi})}}\equiv \sqrt{1+\frac{2\alpha_2 E}{\alpha_1+\alpha_3}}.
\label{kc_axi}
\end{equation}
Perturbations with a wavenumbers $0<k<k_c(E)$ destabilize the base state, since for these wavenumbers $\sigma(k; E)>0$. This is a longwave instability. 
Here I introduced a non-negative parameters $\alpha_1=S\cos{2\psi}, \alpha_2=S\sin{2\psi}$ and a positive parameter 
$\alpha_3=2+S$. At $S=0$ (isotropy) with either $E=0$, or $E\neq 0$, Eqs. (\ref{sigma_axi}) and
(\ref{kc_axi}) reduce to $\sigma(k; 0)=k^2(1-k^2)$ and $k_c(0)=1$, respectively, which characterize a classical Rayleigh-Plateau instability of a free-standing 
solid wire \cite{NM2}. 
In dimensional coordinates this implies that an axisymmetric sinusoidal perturbation induces instability if and only if
its wavelength is longer than the circumference of the undisturbed cylinder.

\begin{figure}[h]
\vspace{-0.2cm}
\centering
\includegraphics[width=3.0in]{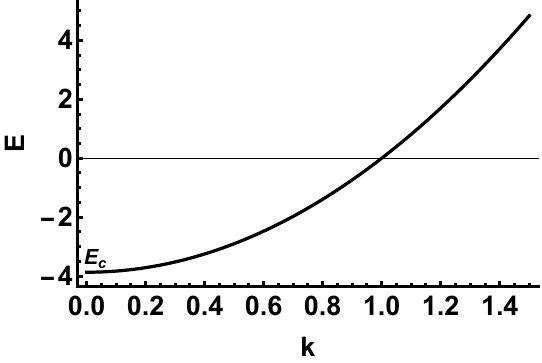}\hspace{0.5cm}
\caption{The neutral stability curve $E(k)$.}
\label{neut}
\end{figure}
\begin{figure}[h]
\vspace{-0.2cm}
\centering
\includegraphics[width=5.25in]{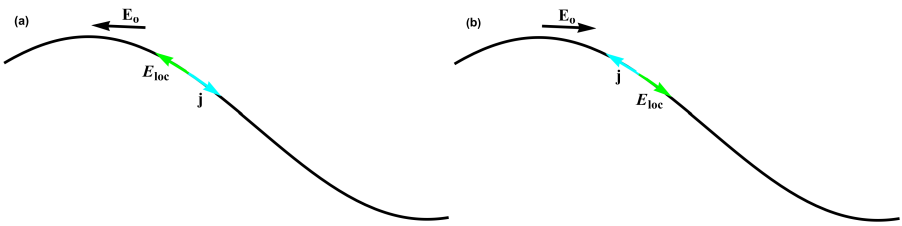}
\caption{(a,b) Sketch of a cross-section of a bump on a wire surface by a plane that is parallel to a wire axis. 
A projection of a constant electric field $E_0$ onto a surface, $E_{loc}$, drives the electromigration adatom flux $j$ 
in the opposite direction, i.e. in the direction of the electron flow $j_e$. This contributes to 
either (a) a bump smoothing (stabilization),
or to (b) a bump growth 
(destabilization). }
\label{Elocal}
\end{figure}

Setting $\sigma=0$ results in the neutral stability curve $E(k)=\frac{\alpha_1+\alpha_3}{2\alpha_2}\left(k^2-1\right)$. 
Fig. \ref{neut} shows this curve for $S=1$ and $\psi=\pi/12$ \cite{MyWire1} (these values are fixed for the remainder of this paper).
The threshold of a longwave instability is $(k,E_c)=(0,\frac{-\left(\alpha_1+\alpha_3\right)}{2\alpha_2})=(0,-3.866)$. Above the neutral stability curve, i.e. for $E>E_c$
the wire is unstable; below the neutral stability curve it is stable.  
For $E>0$ the instability is due to a combined action of two destabilizing factors, the electromigration and a surface diffusion. 
For $E<E_c$ the wire is completely stabilized by electromigration.
For $E_c<E<0$ the stabilizing action of electromigration is weak and the instability due to surface diffusion still emerges. 
Fig. \ref{Elocal} shows why not only the strength of the electric field, but also its direction (determined by a sign of $\Delta V$) matters for a morphological instability. 
For a gold wire, values of the physical parameters can be taken as: $R_0=25$ nm, $\ell=1000$ nm, $Q=20 |e|$ (here $e$ is the electron charge),
$\gamma=1500$ erg$/$cm$^2$, $\Omega=10^{-22}$ cm$^3$. Then at $|E_c|=3.866$ one obtains 
0.01 V for the critical value of the applied voltage.

\underline{\emph{Remark.}} In Ref. \cite{WTLZNSMH}, where electromigration is not considered, a correction to $k_c(0)$ is derived that accounts for 
the finite perturbation amplitude, i.e. the starting assumption $\xi_0\ll 1$ is replaced by $0<\xi_0<1$. The corrected expression reads $k_c(0)=\sqrt{1+\xi_0^2}$.
For the analyses that follow in Sections \ref{Weak} and \ref{MultScale} it is not necessary to likewise correct $k_c$ in Eq. \ref{kc_axi}, since in Sec. \ref{Weak} 
$k \rightarrow 0$ (is tiny) and $k_c$ is irrelevant, and in Sec. \ref{MultScale} $k_c$ is not needed explicitly.

\section{Weakly nonlinear analysis slightly above the instability threshold $(k,E)=(0,E_c)$}
\label{Weak}

\subsection{Derivation of the amplitude equation for the perturbation of the uniform base state $r=1$}

I start by substituting $r= 1+\delta \xi (t,y),\ \delta\ll 1$ in the right-hand side of Eq. (\ref{govPDE}), expanding to second order in $\delta$, 
and formally adding the coefficients of $\delta$  and
$\delta ^2$. This results in:

\begin{eqnarray}
\left(\alpha_3-1\right) \xi_t&=&-\left(E \alpha_2+\frac{1}{2} (\alpha_1+\alpha_3)\right)\xi_{yy} -\frac{1}{2}(\alpha_1+\alpha_3) \xi_{yyyy}+
\frac{1}{2} (\alpha_1+\alpha_3) \xi_y^2-\frac{1}{2} (\alpha_1+\alpha_3) \xi_{yy}^2\nonumber\\
&-&(\alpha_1+\alpha_3) \xi_y \xi_{yyy}+
\left[(E \alpha_2 +\alpha_1+\alpha_3) \xi+\left(2 \alpha_2 -\frac{1}{2} E (5 \alpha_1+\alpha_3)\right) \xi_y+\alpha_2 \xi_{yyy}\right]\xi_{yy} 
+\alpha_2 \xi_y \xi_{yyyy}. \label{phieqn}
\end{eqnarray}
LSA results (\ref{sigma_axi}), (\ref{kc_axi}) may be recovered from Eq. (\ref{phieqn}).

Let $E=E_c+\epsilon  E_1, E_1=\mathcal{O}(1)>0$.
Note that the coefficient of $\xi_{yy}$ changes its sign at $E=E_c$, hence it is proportional to $E-E_c=\epsilon  E_1$, which is $\mathcal{O}(\epsilon)$. 
The balance of the largest nonlinear term $\sim \xi_y^2$ with the linear 
term $\sim\epsilon\xi_{yy}$
is for $\xi(t,y)=\epsilon \phi(t,y), \phi=\mathcal{O}(1)$. 
Also, since the linear instability interval is $0<k<k_c$, where 
$k_c=\sqrt{1+\frac{2\alpha_2 E}{\alpha_1+\alpha_3}}=\sqrt{1-\frac{E}{E_c}}=\sqrt{\frac{-E_1}{E_c}\epsilon}=\mathcal{O}\left(\epsilon^{1/2}\right)$,
the appropriate rescaling of the axial coordinate $y$ is $Y=\epsilon^{1/2}y$.

Thus also let: 
\begin{equation}
\xi=\epsilon \phi;\quad 
T=\epsilon^2 t \Leftrightarrow \partial_t=\epsilon^2 \partial_T,\quad 
Y=\epsilon^{1/2} y \Leftrightarrow \partial_y=\epsilon^{1/2} \partial_Y,
\label{expansions}
\end{equation}
where $T, Y$ are a slow time and a stretched axial variable.

Inserting $E=E_c+\epsilon  E_1$ and expansions (\ref{expansions}) in Eq. (\ref{phieqn}) 
I get:

\begin{equation}
\phi_T \epsilon^3=-\left\{\alpha_2 E_c +\frac{1}{2}\left(\alpha_1+\alpha_3\right)\right\}\phi_{YY}\epsilon^2+\left\{-a \phi_{YY}-b\left[\phi_{YYYY}-\phi_Y^2-\phi\phi_{YY}\right]\right\}\epsilon^3+\mathcal{O}\left(\epsilon^{7/2}\right),
\label{fullexpanded}
\end{equation}
where $a=\frac{\alpha_2 E_1}{\alpha_3-1}=\frac{S E_1 \sin{2\psi}}{1+S}=E_1/4>0$, 
$b=\frac{\alpha_1+\alpha_3}{2\left(\alpha_3-1\right)}=\frac{2+S(1+\cos{2\psi})}{2(1+S)}\approx 0.966$. Noticing that the coefficient of $\phi_{YY}\epsilon^2$ is zero, 
in the lowest order $\epsilon^3$  I finally obtain the amplitude equation
\begin{equation}
\phi_T= -a\; \phi_{YY}-b\left[\phi_{YYYY}-\phi_Y^2-\phi\phi_{YY}\right].
\label{mKS}
\end{equation}
Eq. (\ref{mKS}) can be also written in the form:
\begin{equation}
\phi_T= -a\; \phi_{YY}-b\;\phi_{YYYY}+\frac{1}{2}b\; \partial_{YY}\phi^2.
\label{mKS1}
\end{equation}

Introducing the scalings of $Y$ and $T$:
\begin{equation}
X=\sqrt{\frac{a}{2b}} Y \Leftrightarrow \partial_Y=\sqrt{\frac{a}{2b}} \partial_X,\quad 
\tau=\frac{a^2}{2b} T \Leftrightarrow \partial_T=\frac{a^2}{2b} \partial_\tau
\label{scl}
\end{equation}
eliminates one parameter from Eq. (\ref{mKS1}) and results in equation that has the form of Eq. (9) in Ref. \cite{GB}: 
\begin{equation}
\phi_\tau= -\phi_{XX}-\frac{1}{2} \phi_{XXXX}+c\ \partial_{XX}\phi^2,\quad c=\frac{b}{2a}.
\label{mKS3}
\end{equation}
LSA of Eq. (\ref{mKS3}) about the base state $\phi=1$ gives the 
perturbation growth rate 
$\sigma = (1-2c) k^2 - \frac{k^4}{2}$. 
Thus this base state is longwave unstable at $0<c<1/2\Leftrightarrow 0<b/a<1$, with
the instability cut-off wavenumber $k_c=\sqrt{2(1-2c)}$. For the computation and estimates I'll use $b=1, a=2$, which gives 
$c=1/4$ and $k_c=1$.

Notice that setting $\phi=w_{XX}, c=1$ in Eq. (\ref{mKS3}) and integrating twice with respect to $X$ gives 
$w_\tau= -w_{XX}-w_{XXXX}+w_{XX}^2$ \cite{Bprivate}. This equation was rigorously studied in Ref. \cite{BB} 
(see equation (1D MKS) on p. 377 of that paper, which reads $w_\tau= -w_{XX}-w_{XXXX}+(1-\lambda)w_X^2+\sigma \lambda w_{XX}^2$, 
with $\lambda=\sigma=1$). 
For the 1D MKS equation on a periodic domain these authors proved that 
depending on the stability of the trivial solution, either: (i) if it is unstable,
there exist arbitrarily small initial perturbations that lead to a solution blow-up in a finite time, or (ii) if it is stable, there
exist finite-amplitude initial perturbations that lead to a finite-time blow-up. These singularities exhibit
self-similar structure in $w_{XX}$. Their numerical results indicated that $w$ generically becomes pointwise infinite at a finite time. 
Notice that the limit $\lambda\rightarrow 0$ recovers the Kuramoto-Sivashinsky equation which does not exhibit blow-up in one
dimension. Also, Eq. (\ref{mKS3}) with the $-1$ coefficient of $\phi_{XXXX}$ and $c=-1$ has been studied in Ref. \cite{BW}. For that equation
(the Sivashinsky equation) these authors
found an infinite family of self-similar blow-up solutions, performed their LSA, and identified a unique 
stable blow-up solution. I will refer to Eq. (\ref{mKS3}) also as Sivashinsky equation.

\subsection{Scaling analysis and results}

A convenient scaling analysis of a solution singularity development for Eq. (\ref{mKS3}) 
was performed in Ref. \cite{GB}. 
For the sake of clarity I repeat the major points of this analysis in the next paragraph, using my notation.

Let the singularity forms at location $X =0$ at time $\tau= \tau_s$. 
For $|X|\ll 1$ and $\tau \lesssim \tau_s$, the variable $\phi$ varies rapidly with position and 
$\phi_{XX}$ is negligible compared to $\phi_{XXXX}$. 
Therefore when seeking the behavior of 
$\phi$ close to the singularity, one may consider Eq. (\ref{mKS3}) without the term $-\phi_{XX}$. 
Next, rescaling the distance $\tilde X = c X$ and time $\tilde \tau =c^4 \tau$  gives
\begin{equation}
\phi_{\tilde \tau}= -\frac{1}{2} \phi_{\tilde X \tilde X \tilde X \tilde X}+\frac{1}{c} \partial_{\tilde X \tilde X}\phi^2.
\label{mKS5}
\end{equation}
One seeks similarity solutions to this equation that have the form  
\begin{equation}
\phi(\tilde \tau,\tilde X)=\frac{1}{c\left(\tilde \tau_s - \tilde \tau  \right)^{\gamma_1}}f\left(\frac{\tilde X-\bar X}{\left(\tilde \tau_s - \tilde \tau  \right)^{\gamma_2}}\right),
\label{similform}
\end{equation}
where $\bar X = \bar X(\tilde \tau)$ is the rescaled  position of developing singularity, $\tilde \tau_s=c^4 \tau_s$ 
is the rescaled time of singularity formation, 
$\tilde \tau<\tilde \tau_s$ and $\gamma_1$ and $\gamma_2$ are constant exponents.
Substitution of Eq. (\ref{similform}) in Eq. (\ref{mKS5}) yields
\begin{equation}
\phi(\tau,X)=\frac{1}{c\left(\tau_s - \tau  \right)^{1/2}}f(\eta),
\label{selfsim1}
\end{equation}
where $\eta=(X-x(\tau))/\left(\tau_s - \tau  \right)^{1/4}$ is the similarity variable and $x(\tau)=\bar X(\tilde \tau)/c$.
Eq. (\ref{selfsim1}) signals that
a downward spike in $\phi$ develops that is centered at the point $x(\tau)$; the depth of the spike increases as 
$\left(\tau_s - \tau  \right)^{-1/2}$ and its width decreases as $\left(\tau_s - \tau  \right)^{1/4}$ as 
$\tau \rightarrow \tau_s$ from below. 
The unscaled position of the spike's 
center moves with velocity $x_\tau\sim \left(\tau_s-\tau\right)^{-3/4}$. 

\begin{figure}[h]
\vspace{-0.2cm}
\centering
\includegraphics[width=5.75in]{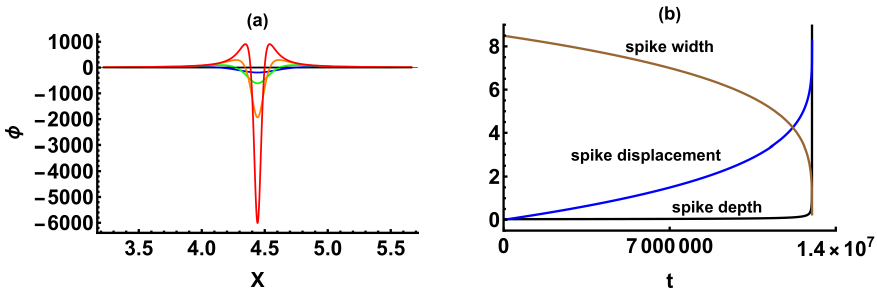}
\caption{(a) Spike formation via computation of Eq. (\ref{mKS3}) with $c=1/4$. 
The initial condition is a small Gaussian-shaped perturbation of the base state $\phi=1$ on the interval 
$0\le X\le 2\sqrt{2}\pi$, where $2\sqrt{2}\pi$ is the most dangerous instability wavelength (the one at which the growth rate $\omega$ 
is the maximum value). Boundary conditions are periodic.
The last shown profile corresponds to $\tau_s=43.35$. Computation was done on a fixed grid in $X$ using 3000 grid points, 
a fourth-order finite difference discretization formulas in $X$ and a backward difference formulas of variable order in $\tau$.
(b) Evolution of the spike dimensions (Eqs. (\ref{spd})-(\ref{spdisp})) at $\Delta E=0.01, a=2, b=1, t_s=1.3\times 10^7$. Proportionality constants 
$c_{1,2,3}=1$ are chosen to facilitate plotting three functions on the same plot.
}
\label{spk}
\end{figure}

In terms of the original variables used in Eq. (\ref{phieqn}):
\begin{equation}
\mbox{spike depth} = \frac{c_1}{\Delta E \sqrt{b\left(t_s-t\right)}},
\label{spd}
\end{equation}
\begin{equation}
\mbox{spike width} = c_2 \sqrt{a\Delta E} \left(\frac{t_s-t}{b}\right)^{1/4},
\label{spw}
\end{equation}
\begin{equation}
\mbox{spike center displacement from}\ y=0 =  \frac{c_3\sqrt{a\Delta E}}{b^{1/4}}\left[t_s^{1/4}-\left(t_s-t\right)^{1/4}\right],
\label{spdisp}
\end{equation}
where $\Delta E = E-E_c$ and the time of a singularity formation $t_s$ and the proportionality constants $c_{1,2,3}$ can't be determined from the 
scaling analysis. Obviously, one expects that $t_s=t_s(\Delta E)$, thus Eqs. (\ref{spd})-(\ref{spdisp}) show the dependence of the spike parameters
on time, but not yet on $\Delta E$.  In order to determine that dependence, first, 
I use Eqs. (\ref{expansions}), (\ref{scl}), $\epsilon=\Delta E/E_1$, and $a=E_1/4$ to find $t_s$: 
\begin{equation}
t_s=\frac{2b \tau_s}{a^2 \epsilon^2}=\frac{32 b\tau_s}{\Delta E^2}.
\label{ts}
\end{equation}
Next, I compute Eq. (\ref{mKS3}), see Fig. \ref{spk}(a), which gives $\tau_s\approx 40$. Using this value, 
$b=1$ and $\Delta E =0.01$ gives $t_s\approx 1.3\times 10^7$. Evolution of the spike dimensions at these (fixed) 
values of $\Delta E$ and $t_s$ is shown in 
Fig. \ref{spk}(b). It is seen that in comparison to the singularity of the width and the displacement, the singularity of the depth develops abruptly. 

Next, inserting Eq. (\ref{ts}) in Eqs. (\ref{spd})-(\ref{spdisp}) gives the following forms for the 
dependence of the spike parameters on $\Delta E$ and $t$:
\begin{equation}
\mbox{spike depth} = \frac{c_1}{b F(t, \Delta E)^2},
\label{spd1}
\end{equation}
\begin{equation}
\mbox{spike width} = c_2 \sqrt{a}F(t, \Delta E),
\label{spw1}
\end{equation}
\begin{equation}
\mbox{spike center displacement from}\ y=0 
= c_3\sqrt{a}\left[2\left(2\tau_s\right)^{1/4}-
F(t, \Delta E)\right],
\label{spdisp1}
\end{equation}
where
\begin{equation}
F(t, \Delta E) = \left(32\tau_s-\frac{\Delta E^2 t}{b}\right)^{1/4}.
\label{FF}
\end{equation}

\begin{figure}[h]
\vspace{-0.2cm}
\centering
\includegraphics[width=5.75in]{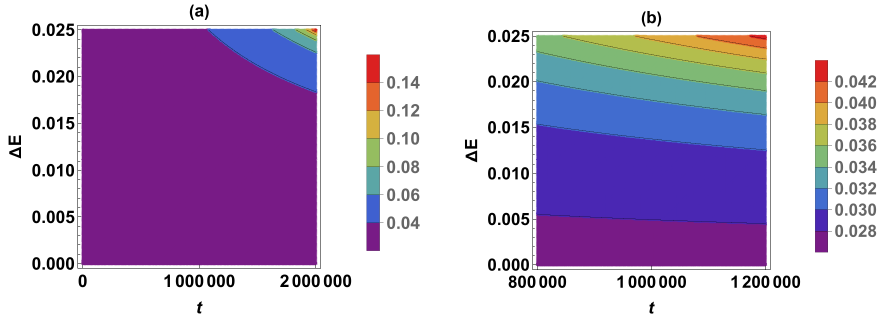}
\caption{Spike depth, as given by Eq. (\ref{spd1}). $b=1, \tau_s=40, c_1=1$. (b) is the zoom view of (a) in the middle of the time interval. 
}
\label{spkdth}
\end{figure}

Figure \ref{spkdth} shows the major spike dimension, its depth, as the contour plot of the two-variable function (\ref{spd1}). 
Spike depth increases as the time and $\Delta E$ increase. Moreover, unless the time is close to a blow-up and $\Delta E$ is not very small, the depth is more sensitive 
to $\Delta E$ change than to $t$ change. This is the signature of Eq. (\ref{FF}), where the exponent of $t$ 
is one, and the exponent of $\Delta E$ is two. Same is true for the spike width and displacement.
%
Also it is useful to expand Eq. (\ref{spd1}) for small $\Delta E$ and $t$:
\begin{equation}
\mbox{spike depth} = 
\frac{c_1}{4b \sqrt{2\tau_s}}+\frac{c_1}{256\sqrt{2} b^2  \tau_s^{3/2}}\Delta E^2 t+O\left(\Delta E^4 t^2\right).
\label{spd2}
\end{equation}
This shows that the spike depth grows quadratically in $\Delta E$ and only linear in $t$.

\subsection{Steady-state blow-up profile of $\phi$}

It is interesting to attempt finding a constant (steady-state) solution of Eq. (\ref{mKS3}) or Eq. (\ref{mKS5}). In the case of Eq. (\ref{mKS3}), 
setting $\phi_\tau=0$ and integrating twice in $X$ gives:
\begin{equation}
\phi_{XX}+2\phi-2c \phi^2=k_1 X+k_2,\label{osc}
\end{equation} 
where $k_1, k_2$ are the constants of integration. Eq. (\ref{osc}) is the equation of a forced nonlinear oscillator,
where $X$ is the ``time". I did not succeed in finding a solution of Eq. (\ref{osc}) that would be symmetric, $\phi(-X)=\phi(X)$ and singular at $X=0$. 
In the case of Eq. (\ref{mKS5}), one obtains a steady-state equation
\begin{equation}
\phi_{\tilde X \tilde X}-\frac{2}{c}\phi^2=k_1 \tilde X+k_2.\label{osc1}
\end{equation} 
At $k_1=0$ and arbitrary $c>0$, Mathematica\textsuperscript{\tiny\textregistered} finds the general solution of Eq. (\ref{osc1}) in terms of Weierstrass elliptic function $W(\tilde X,c,k_2,l_1,l_2)$, where
$l_1, l_2$ are arbitrary constants. Remembering that Eqs. (\ref{mKS5}) and (\ref{osc1}) hold in the vicinity of a spike tip at $\tilde X=0$, 
a particular symmetric solution that blows up at $\tilde X=0$ and is zero at some distance away from $\tilde X=0$
can be guessed by taking $l_1=0$ and some positive values for $k_2$ and $l_2$. A power series expansion about $\tilde X=0$ with just a 
handful of terms very accurately approximates such solution in the vicinity of $\tilde X=0$. For instance, at $k_2=10, c=1/4, l_2=2$:
\begin{equation}
\phi=\frac{-3}{4} \tilde X^{-2} + \tilde X^2 - \frac{2}{21} \tilde X^4+\mathcal{O}\left(\tilde X^6\right). \label{osc1sol}
\end{equation} 
\begin{figure}[h]
\vspace{-0.2cm}
\centering
\includegraphics[width=3.0in]{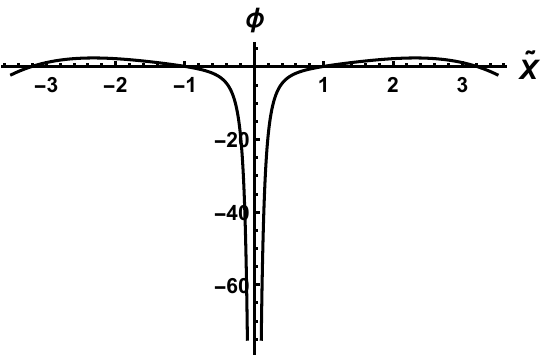}
\caption{Plot of Eq. (\ref{osc1sol}).  
}
\label{spkstd}
\end{figure}

Instead of finding a steady-state amplitude spike by the direct integration of the amplitude equation, Bernoff \& Bertozzi \cite{BB}
computed a self-similar blow-up profile for $f(\eta)$, see Eq. (\ref{selfsim1}).  
Substitute Eq. (\ref{selfsim1}) in Eq. (\ref{mKS5}), transform to a singular (at $\eta=0$) third-order equation \cite{BB}, 
drop the time derivative, and require the symmetry $f(-\eta)=f(\eta)$. This yields:
\begin{eqnarray}
\frac{1}{2}\eta f_{\eta\eta\eta}-f_{\eta\eta}-\eta\left(f^2\right)_\eta+\frac{1}{4}\eta^2 f+f^2&=&f^2(0)-f_{\eta\eta}(0),\label{ssim1}\\
f(-\infty)=f(\infty)=f_\eta(0)&=&0.\label{ssim2}
\end{eqnarray}
Bernoff \& Bertozzi computed the solution of the above boundary value problem using a shooting method, except the coefficient 
of $\eta f_{\eta\eta\eta}$ in their equation is equal to one. That solution is shown in Fig. 4 of their paper. 
The shape is visually very similar to the spike in Fig. \ref{spk}(a), where the axes are $\eta$ and $f$ and the profile is shifted to 
the left, so that the symmetry axis is $\eta=0$. The smooth minimum of the profile is at $(\eta,f)=(0,-2.169)$ and $f_{\eta \eta}(0)=2.694$ \cite{BB}.
Having a coefficient 1/2 of $\eta f_{\eta\eta\eta}$ in Eq. (\ref{ssim1}) of course would not change that shape in any significant way.

\subsection{Physical estimates}

For a gold wire, using $R_0=25$ nm, $T=800^\circ$C, $D=5\times 10^{-7}$ cm$^2/$s, 
$\gamma=1500$ erg$/$cm$^2$, $\Omega=10^{-22}$ cm$^3$, $\nu=5\times 10^{14}$ cm$^{-2}$, Eq. (\ref{ts}) gives $t_s=5$ hr.
At that temperature, this is the 
typical physical time of a singularity formation just above the instability threshold ($\Delta E =0.01$). For comparison, at $T=600^\circ$C 
this time increases to 50 hr. due to the ten-fold decrease of the diffusivity, and at $T=800^\circ$C and $\Delta E =0.1$, 
i.e. at the limit of a validity of a weakly nonlinear theory, this time decreases to 3 min. In any case, the pinch-off time predicted 
by Eq. (\ref{ts}) is significantly larger than the one in experiment; for example, Table 1 in Ref. \cite{BBSN} suggests that 
a silver nanowire of 25 nm radius would pinch-off in about 10 min at $T=300^\circ$C. 
This difference may be the sign that an advanced model that includes a wire heating due to a current crowding is warranted. On the other hand,
it should be remembered that the amplitude equation (\ref{mKS3}) is derived assuming a tiny supercriticality $\Delta E$, i.e. 
the system is just above the instability threshold (formally, it is at a distance $\epsilon \ll 1$ above a threshold at all times), 
thus the amplitude is expected to grow very slow. A weakly nonlinear analysis of any advanced model would have the same limitation on the distance 
from the instability threshold simply due to a nature of such analysis. Next, one may try to estimate the separation between two parts 
of a wire upon a pinchoff. In Fig. \ref{spk}(a), the width of the spike at the half minimum $\phi\approx -3000$, 
corresponding to $r\approx 1/2$, is $\Delta X\approx 0.1$. A computation that uses an adaptive mesh refinement at the tip of the spike would be able to follow
the spike deepening somewhat longer, thus the half minimum level would shift downward and one may take, say, $\Delta X\approx 0.025$.  Then our scalings imply the separation at $r\approx 1/2$, 
$\Delta y\approx R_0\sqrt{\frac{10}{\Delta E}} \Delta X=18$ nm at $\Delta E=0.01, R_0=25$ nm, or around 63 atomic diameters of 
the gold atom. $\Delta y$ reduces to 6 nm at $\Delta E=0.1$.


\section{A weakly nonlinear multi-scale analysis of the primary modes of instability}
\label{MultScale}

\subsection{Multiple scales expansion}
As was already noted in Introduction, due to the nonlinearity of a governing PDE (\ref{govPDE}) very shortly after the initiation of a morphological instability 
the initial sinusoidal, single-harmonic form of a 
solution (the initial condition) ceases to exist, and the solution is represented by a Fourier series.
Without loss of generality, I will account only for a handful of major terms in that series, i.e. the sinusoidal terms having a wavenumbers that are a small integer multiples 
of a wavenumber of the primary mode (the sinusoidal initial condition/perturbation) \cite{WTLZNSMH}, and derive their time-dependent amplitudes. 
The primary mode
has a finite wavenumber $0<k<k_c$.

I start with the following ansatz for the perturbation of the base state $r=1$: 
\begin{equation}
r= 1+\epsilon r_1\left(t_0,t_1,t_2; E\right)\cos{k y}+\epsilon r_4\left(t_0,t_1,t_2; E\right)\sin{k y}+
\epsilon^2 r_2\left(t_0,t_1,t_2; E\right)\cos{2k y}+\epsilon^2 r_3\left(t_0,t_1,t_2; E\right)\sin{2k y},\ \epsilon\ll 1
\label{rexpansion}
\end{equation}
where $t_0$ is the fast time and $t_1, t_2$ are the slow times. Thus the time derivative reads:
\begin{equation}
\frac{\partial}{\partial t}=\frac{\partial}{\partial t_0}+\epsilon \frac{\partial}{\partial t_1}+\epsilon^2\frac{\partial}{\partial t_2}.
\label{ddtexpansion}
\end{equation} 

The initial conditions are chosen as:
\begin{equation}
r_1\left(0,t_1,t_2; E\right)=\xi_0,\; 0<\xi_0<1;\quad r_{2,3,4}\left(0,t_1,t_2; E\right)=0.
\label{inic}
\end{equation}

Now I proceed according to this plan: substitute the expansions (\ref{rexpansion}) and (\ref{ddtexpansion}) in Eq. (\ref{govPDE}), collect the terms of the same powers of 
$\epsilon$ on each of the two sides of the equation and equate them; next, for each of the resulting statements, collect the terms proportional 
to $\cos{k y}$, $\sin{k y}$, $\cos{2k y}$, and $\sin{2k y}$ on each of the two sides of a statement and equate, separately, the coefficients 
of these harmonics.   

At the order $\epsilon$ I get:
\begin{equation}
\frac{\partial r_1}{\partial t_0}\cos{k y}+\frac{\partial r_4}{\partial t_0}\sin{k y}=\sigma(k; E) r_1 \cos{k y} + \sigma(k; E) r_4 \sin{k y},
\label{ordereps}
\end{equation}
thus
\begin{equation}
\frac{\partial r_{1,4}}{\partial t_0}=\sigma(k; E) r_{1,4},
\label{r14equations}
\end{equation}
where the growth rate $\sigma(k; E)$ is given by Eq. (\ref{sigma_axi}). 

At the order $\epsilon^2$ I get:
\begin{equation}
\frac{\partial r_1}{\partial t_1}\cos{k y}+\frac{\partial r_4}{\partial t_1}\sin{k y}+\frac{\partial r_2}{\partial t_0}\cos{2k y}+
\frac{\partial r_3}{\partial t_0}\sin{2k y}=f_1\left(k,r_1,r_4; E\right)+f_2\left(k,r_1,r_4,r_2; E\right) \cos{2k y} + f_3\left(k,r_1,r_4,r_3; E\right) \sin{2k y},
\label{orderepssq}
\end{equation}
where $f_{1,2,3}$ are certain complicated functions. 

\subsection{Solutions for $r_1$ and $r_4$}
\begin{equation}
\frac{\partial r_{1,4}}{\partial t_1}=0 \rightarrow r_{1,4}=r_{1,4}\left(t_0,t_2; E\right). 
\label{r1r4_1}
\end{equation}
Applying the initial conditions (\ref{inic}) and accounting for Eq. (\ref{r1r4_1}) I get from the linear ODE (\ref{r14equations}):
\begin{equation}
r_1 = r_1\left(t_0; E\right)=
\xi_0 e^{\sigma(k; E) t_0},\quad r_4=0.
\label{r1r4sols}
\end{equation}
Here $\sigma(k; E)>0$, since the case of unstable primary mode is of interest. It can be observed now that the LSA result of Sec. \ref{LinStab} for the primary mode 
has been obtained at the linear stage (at order $\epsilon$) of a multi-scale expansion. 

\subsection{Solutions for $r_2$ and $r_3$}
Substituting $r_4=0$ in $f_2$ and $f_3$, for $r_2$ and  $r_3$ from Eq. (\ref{orderepssq}) 
one obtains the linear ODEs:
\begin{equation}
\frac{\partial r_{2,3}}{\partial t_0}=f_{2,3}\left(k,r_1,r_{2,3}; E\right)=p(k; E) r_{2,3}+ q_{2,3}(k; E)r_1\left(t_0; E\right)^2,\label{r23equations}
\end{equation}
where
\begin{equation}
p(k; E)=2\frac{2+S(1+\cos{2 \psi})}{1+S}k^2\left(1-4k^2\right)+4E \frac{S\sin{2 \psi}}{1+S}k^2=\sigma(2k; E),
\end{equation}
\begin{equation}
q_2(k; E)=-3\frac{2+S(1+\cos{2 \psi})}{4(1+S)}k^2\left(1+k^2\right)-E \frac{S\sin{2 \psi}}{2(1+S)}k^2,
\end{equation}
\begin{equation}
q_3(k; E)=\frac{S\sin{2 \psi}}{1+S}k^3\left(1-k^2\right)+E \frac{S^2(\cos{4 \psi}-1)-S(6\cos{2\psi}+10)-12}{4(1+S)}k^3.
\end{equation}
Substituting $r_1\left(t_0; E\right)$ from Eq. (\ref{r1r4sols}) and taking into account the initial conditions, the solutions of Eqs. (\ref{r23equations}) read:
\begin{equation}
r_{2,3}=r_{2,3}\left(t_0; E\right)=\frac{q_{2,3}(k; E)\xi_0^2}{\sigma(2k; E)-2\sigma(k; E)}\left[\exp{\left(\sigma(2k; E)t_0\right)}-\exp{\left(2\sigma(k; E)t_0\right)}\right].
\label{r2r3sols}
\end{equation}

\subsection{Compatibility condition}
Finally, for consistency of Eq. (\ref{orderepssq}) one needs $f_1\left(k,r_1; E\right)=0$. 
This gives the constraint $k^2\left(k^2-k_c^2\right)r_1\left(t_0; E\right)^2=0$, where $k_c$ is seen in Eq. (\ref{kc_axi}). 
Since $r_1\left(t_0; E\right)\neq 0$, either $k=0$, or $k=k_c$. 
I proceed below under assumption that the constraint  
holds approximately at $k=\delta_1$ 
and $k=k_c-\delta_2$, where $\delta_{1,2}$ are a positive constants that quantify the deviation from these values into the instability interval $0<k<k_c$.
In particular, $k=\delta_1$ is the most dangerous wavenumber in LSA when $\delta_1=k_c/\sqrt{2}$.

\subsection{Results}

\begin{figure}[h]
\vspace{-0.2cm}
\centering
\includegraphics[width=5.75in]{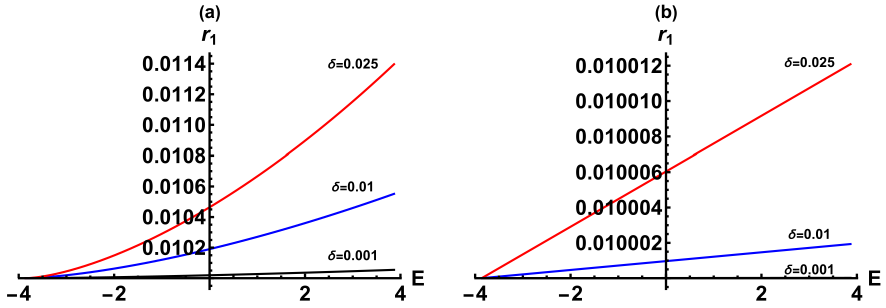}
\caption{The amplitude of $\cos{k y}$ term in the ansatz (\ref{rexpansion}), given by Eq. (\ref{r1r4sols}), at $t_0=1$, $\xi_0=0.01$ for 
$E_c+0.01\le E\le -E_c$, where $E_c=-3.866$.  (a) Near the short-wave instability cut-off; (b) Near the long-wave instability 
threshold.
}
\label{r1_vs_E}
\end{figure}

Figure \ref{r1_vs_E} shows $r_1$ as a function of the electric field at fixed $t_0$ and at various distances $\delta$ from the longwave instability threshold 
$k=0$ and from the short-wave instability cut-off $k=k_c$. For this, I took $\delta_1=\delta_2=\delta$ and substituted $k=\delta$ and $k=k_c-\delta$ in Eq. (\ref{r1r4sols}), where 
$\sigma(k; E)$ and $k_c(E)$ are given by Eqs. (\ref{sigma_axi}), (\ref{kc_axi}). In the former case (Fig. \ref{r1_vs_E}(b)) one can see that 
$r_1\sim c_1\exp{(c_2 E)}, c_1, c_2=const.$, and 
values of these constants can be easily calculated. However, for the particular numerical values of the parameters and for the plotting interval 
these exponential curves are well approximated by the straight lines.  In the latter case 
(Fig. \ref{r1_vs_E}(a)) the analytical dependence on $E$ is complicated, but the curve data can be fitted: $\delta=0.001$: $r_1=0.00997 + 0.000048\exp{(0.145717 E)}$; 
$\delta=0.01$: $r_1=0.00973 + 0.000458 \exp{(0.153945 E)}$; $\delta=0.025$: $r_1=0.00941 + 0.001056 \exp{(0.16734 E)}$.

\begin{figure}[h]
\vspace{-0.2cm}
\centering
\includegraphics[width=5.25in]{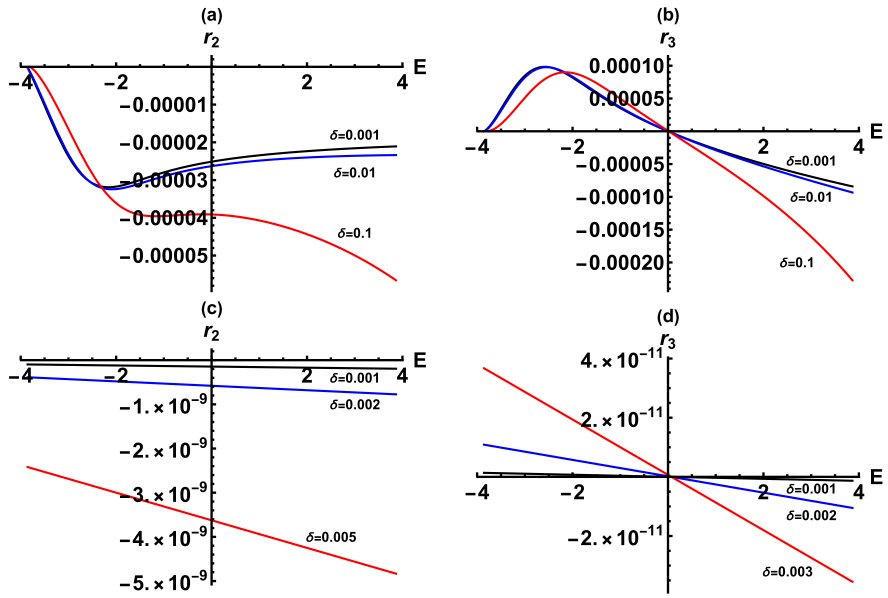}
\caption{The amplitudes of $\cos{2k y}$ and $\sin{2k y}$ terms in the ansatz (\ref{rexpansion}), given by Eqs. (\ref{r2r3sols}), 
at $t_0=1$,  $\xi_0=0.01$ for $E_c+0.01\le E\le -E_c$, where $E_c=-3.866$.  (a,b) Near the short-wave instability cut-off; (c,d) Near the long-wave instability 
threshold.
}
\label{r23_vs_E}
\end{figure}

A close look at $r_2(E)$ and $r_3(E)$ curves (Fig. \ref{r23_vs_E}) and the comparison to $r_1(E)$ in Fig. \ref{r1_vs_E} shows, first, that $r_1$
is several orders of magnitude larger than $r_{2,3}$. Thus the primary mode $\cos{ky}$ is dominant in the spectrum, 
as expected \cite{WTLZNSMH}.  Next, the absolute values of $r_{2}$ are two orders of magnitude larger than the absolute values of $r_{3}$ 
near the longwave instability threshold, while the converse is true near the short-wave instability cut-off. 
Therefore $\cos{2ky}$ and $\sin{2ky}$ are the next two fast growing modes in the spectrum. 

\begin{figure}[h]
\vspace{-0.2cm}
\centering
\includegraphics[width=5.75in]{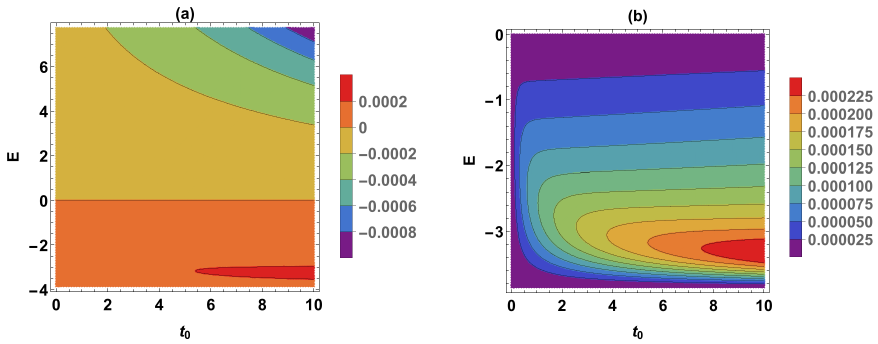}
\caption{The amplitude of $\sin{2k y}$ term in the ansatz (\ref{rexpansion}), given by Eqs. (\ref{r2r3sols}), at $\xi_0=0.01$, near the short-wave 
instability cut-off. (b) is the zoom view of the bottom region in (a). 
}
\label{r3_vs_E_t0}
\end{figure}

The largest of $r_{2,3}$ amplitudes, $r_3$, is plotted in Fig. \ref{r3_vs_E_t0} vs. $t_0$ and $E$ near the short-wave instability cut-off. 
Cross-secting Fig. \ref{r3_vs_E_t0}(a) by a vertical cut shows that at any fixed $t_0$ the amplitude dependence on $E$ is roughly as seen in 
Fig. \ref{r23_vs_E}(b). Moreover, a (positive) maximum of the amplitude increases with $t_0$, as seen in 
Fig. \ref{r3_vs_E_t0}(b).
%


\section{Summary}
\label{Summ}

Provided that the electric field slightly exceeds a threshold value that is necessary for initiation of a morphological 
instability, the weakly nonlinear analysis that I carried out shows for the first time the electric field-and-time 
dependence of the dimensions 
of the focusing self-similar perturbation's amplitude profile approaching a blow-up. That amplitude singularity marks a wire 
pinch-off and its breakup into a chain of nanoparticles. 
In particular, this analysis 
shows that the amplitude initially grows quadratically in the deviation from a threshold value of the electric field. 
For the initial surface perturbation of the form $\xi_0\cos{ky},\; 0<\xi_0<1$ (where $y$ is the axial variable) a separate, multi-scale analysis of 
a weakly nonlinear phase of the instability shows
that in that regime the fastest growing instability modes are $\cos{ky}$, $\cos{2ky}$, and $\sin{2ky}$, and for these modes I found the explicit dependence of their amplitudes on time 
and the strength of the applied electric field.

\section{Acknowledgments}
\label{Ack}

The author thanks Alexander A. Nepomnyashchy and Mark R. Bradley for very useful suggestions and discussions regarding the analysis in Sec. \ref{Weak}. 
Mark R. Bradley is also acknowledged for bringing his paper \cite{GB} to author's attention and for pointing to Ref. \cite{BB}.

\end{document}